# Advancing privacy in learning analytics using differential privacy

QINYI LIU, Centre for the Science of Learning & Technology (SLATE), University of Bergen, Norway

RONAS SHAKYA, Centre for the Science of Learning & Technology (SLATE), University of Bergen, Norway

MOHAMMAD KHALIL, Centre for the Science of Learning & Technology (SLATE), University of Bergen, Norway

JELENA JOVANOVIC, Faculty of Organisational Sciences, University of Belgrade; Centre for the Science of Learning & Technology (SLATE), University of Bergen, Norway

This paper addresses the challenge of balancing learner data privacy with the use of data in learning analytics (LA) by proposing a novel framework by applying Differential Privacy (DP). The need for more robust privacy protection keeps increasing, driven by evolving legal regulations and heightened privacy concerns, as well as traditional anonymization methods being insufficient for the complexities of educational data. To address this, we introduce the first DP framework specifically designed for LA and provide practical guidance for its implementation. We demonstrate the use of this framework through a LA usage scenario and validate DP in safeguarding data privacy against potential attacks through an experiment on a well-known LA dataset. Additionally, we explore the trade-offs between data privacy and utility across various DP settings. Our work contributes to the field of LA by offering a practical DP framework that can support researchers and practitioners in adopting DP in their works.

## 1 Introduction

Data collected about learners and learning processes is a cornerstone for advancing learning analytics (LA) research and practice. However, collecting and analysing such data is prone to serious data privacy violations, as malicious actors could potentially misuse it to infringe on the learners' privacy. For instance, the 2023 MOVEit Transfer hack affected over 800 educational institutions, leaking nearly 1.7 million records[4]. This attack was part of a broader trend of targeting organisations with large datasets, including schools and universities. Meanwhile, stricter legal regulations are being introduced worldwide, such as the General Data Protection Regulation (GDPR), Data Governance Act, and California Privacy Rights Act (CPRA) [44]. These laws have shifted the protection of learner privacy from a non-functional requirement and moral obligation to a mandatory regulation. Against this backdrop, rising privacy risks have been

Authors' Contact Information: Qinyi Liu, Centre for the Science of Learning & Technology (SLATE), University of Bergen, Bergen, Norway, qinyi.liu@uib.no; Ronas Shakya, Centre for the Science of Learning & Technology (SLATE), University of Bergen, Bergen, Norway, ronas.shakya@uib.no; Mohammad Khalil, Centre for the Science of Learning & Technology (SLATE), University of Bergen, Bergen, Norway, mohammad.khalil@uib.no; Jelena Jovanovic, Faculty of Organisational Sciences, University of Belgrade; Centre for the Science of Learning & Technology (SLATE), University of Bergen, Bergen, Norway, jelena.jovanovic@uib.no.

recognised as an important concern in the LA community. As privacy protection has become a legal requirement and social consensus, if not appropriately handled, it may prevent LA from reaching its full expected potential [39][30]. This is because learners and other involved stakeholders (e.g., teachers, researchers, and practitioners) may become increasingly hesitant towards sharing their data, due to privacy concerns, resulting in LA struggling to get access to comprehensive, high quality, and large amounts of data for analysis.

The evolving demand for data privacy has also driven the development of privacy-preserving techniques. Over the years, researchers have proposed various methods to protect privacy via algorithms, such as k-anonymity, l-diversity, m-invariance, and t-closeness [43]. However, since these methods have been jailbroken, more advanced privacy-preserving techniques have become necessary [10][36]. Differential privacy (DP) is an advanced privacy-preserving technique that introduces controlled noise into the data analysis process to balance the trade-off between data utility and individual privacy [10]. The advantages of DP are significant: DP is robust against composition attacks[1]. It can also defend against many other attacks on sensitive data, such as bias attacks[2][3][32]. Furthermore, DP may scale to large datasets and complex queries [32]. Due to these advantages, DP has become a widely used privacy protection approach in statistical data analysis and machine learning over the past decade [6]. DP has also been applied in various fields, including IoT, healthcare, and census data, and has been well adopted in industry [6].

In the context of LA, responding to many privacy measures is more of a framework or policy recommendation, with limited practical application evidence [40][30]. Additionally, prior research has shown that conventional data anonymization and de-identification methods are incapable of addressing the complexity and diversity of learning data [23][30]. Motivated by these findings and the compelling recent calls to expand the horizons of LA by integrating insights from other fields, the current paper presents an empirical study to implement and evaluate the application of DP in LA approaches that rely on machine learning. Our contributions are as follows:

· Introduction of the first DP framework for education and learning analytics: We propose the first framework for applying Differential Privacy (DP) to the educational sector, providing practitioners with guidance on implementing DP in LA contexts.

· Proof-of-concept: We empirically demonstrate the feasibility of our framework using a well-known LA dataset.

· Privacy-utility trade-off analysis: Our work includes a comprehensive analysis of the results and showcases how different DP configurations impact both the privacy guarantees and the utility of the data for LA tasks.

## 2 Preliminary of concepts

### 2.1 Basics of differential privacy

The concept of DP, proposed by [11], was introduced in cryptography to keep information secure and private, even if some details are known. [10] demonstrated that any method of accessing a database with sensitive data would automatically entail a non-zero risk of leakage. Such a risk is not limited to individuals whose data is included in the database; even those who have not contributed their data may be exposed, as the analysis of similar or related data within the database can inadvertently reveal information about them [13]. For instance, consider an individual, Alice, who is trying to make a decision to participate in a dataset. We refer to the version of the dataset that includes Alice's records as D0, while the version without her records is D1. When D0 and D1 differ only by one record, such datasets are referred to as neighbouring datasets. DP attempts to limit the difference in the response probabilities of a query[4] applied to neighbouring datasets by adding noise. The following definition formalises this concept.

---

[1] a type of attack when attackers combine multiple independently released anonymized datasets to uncover sensitive information of an individual.

[2] bias attack generally refers to a situation where a malicious actor manipulates a system's output by exploiting biases in the model or the data it was trained on.

[3] a similarity attack occurs when an attacker tries to infer information about an individual's data by comparing the output of a model or a dataset in response to different queries. The attacker exploits the similarity of the responses to find patterns or identify individuals based on their input data.

**Definition:** $\epsilon$-Differential Privacy ($\epsilon$-DP)

A random algorithm $M$ satisfies $\epsilon$-DP if for any two neighbouring datasets $D$ and $D^{'}$ (differing by the data of exactly one individual), and for all possible outcomes of the algorithm $T \subseteq \text{Range}(M)$:

$$\Pr[M(D) \in T] \leq e^{\epsilon} \Pr[M(D^{'}) \in T] \tag{1}$$

In this definition, $Pr$ denotes the probability that the output of the mechanism $M$, given dataset $D$, falls within the subset $T$. $M$ is a randomized mechanism that ensures privacy by adding controlled noise to the data, minimizing the impact of any single individual's data modification on the algorithm's output. $T$ represents a subset of possible outcomes of $M$, capturing specific results that the algorithm might produce. $e^{\epsilon}$ is a multiplicative factor, with $\epsilon$ being a small positive parameter that quantifies the level of privacy. It is also worth mentioning that while this paper relies on the most basic DP definition (the $\epsilon$-DP definition introduced above), there are other DP definition variants [12].

**Privacy Budget**. The privacy budget, represented by epsilon ($\epsilon$), is a key concept in DP. It controls the balance between privacy protection and data utility. A smaller $\epsilon$ means that the differences in the results produced by an algorithm $M$ on neighbouring datasets are minimal. This makes it harder for an attacker to determine whether a specific individual's data is in the dataset, thereby providing stronger privacy protection. However, the increased privacy reduces the utility of the data, as the accuracy of the results is also reduced. On the other hand, a larger $\epsilon$ allows for more noticeable differences between neighbouring datasets, improving data utility but weakening privacy protection [18].

The value of $\epsilon$ typically depends on the specific context of use. In DP for machine learning models, it is widely accepted that when $\epsilon \leq 1$, the system provides strong formal privacy guarantees. Values of $\epsilon$ between 1 and 10 offer a reasonable balance between privacy and utility [38].

**Sensitivity** Sensitivity helps determine the amount of noise needed to protect privacy. Specifically, sensitivity indicates the maximum change in the query output when an element in the dataset changes. The greater the sensitivity, the more sensitive the query is to individual data points in the dataset. For example, if we want to calculate the average of two datasets separately, and the sensitivity of this average query is 1, then, if we add or delete one of the data records in one of the two datasets, the maximum change in the average query is 1. The sensitivity of query $Q$ is denoted by $S$ as defined in (2) [17].

$$S = \max \|Q(D) - Q(D^{'})\| \tag{2}$$

Where, $Q$ is the query function, and $D$ and $D^{'}$ are neighbouring datasets. Note that this $S$ is distinct from the set $S$ used earlier in the definition of differential privacy. $S$ here represents the sensitivity of the query function $Q$.

**Noise** The specific meaningless information added to each data record is referred to as noise. By adding noise, it is ensured that the presence or absence of a single record does not significantly affect the query results, thus protecting privacy. The privacy budget epsilon ($\epsilon$) and sensitivity ($S$) together determine the type of noise in DP [38]. Common types of noise include Gaussian noise, and Laplace noise [38].

## 2.2 Different approaches to data protection with differential privacy in machine learning

Since this paper focuses on privacy issues in LA approaches that rely on machine learning, we introduce distinct approaches for incorporating DP in a machine learning process.

---

[4] In differential privacy, a "query" refers to performing a computation or operation on a database or dataset to obtain specific information or results. A query can be any type of data operation or analysis task, taking the dataset as input and returning some form of output. Differential privacy ensures that the outputs of these queries do not disclose information about individuals in the data to a significant extent. Common queries include Counting Queries (calculating the number of records that meet a certain condition), among others. The training and inference processes of machine learning are also considered types of queries.

*1. Adding DP at the input/data level:* If the input data for a machine learning algorithm is protected by DP, then any model trained using this data will also be differentially private[5], and the model's output will maintain the same privacy protection. This type of DP provides the broadest privacy coverage for stakeholders [38]. However, adding DP at the input stage is highly challenging. Currently, there are two main approaches: local differential privacy (LDP) and synthetic private data generation. LDP is a well-known privacy model for distributed architectures that aims to provide privacy guarantees for each user while collecting and analysing data. However, the noise introduced by LDP is usually significant, which can severely impact the utility of the model. As a result, the mainstream approach in the DP field is to use relaxed forms of LDP, which involve modifying the strict privacy guarantees of LDP to achieve a better balance between privacy protection and data utility. However, [19] demonstrated that even relaxed forms of LDP increase the risk of privacy leakage. On the other hand, synthetic private data generation does not add noise to individual examples in the dataset but seeks to generate entirely private synthetic examples that can be publicly shared. To generate such synthetic data, a probabilistic model that describes the underlying data distribution needs to be created and then sampled. The quality of this model is crucial for the utility of the underlying synthetic data [38].

*2. Adding DP during the machine learning training process:* This is currently the most common method for obtaining differentially private machine learning models [38]. The underlying logic of this approach is that although the input data is sensitive, if the model training algorithm is differentially private, the resulting model and its output will also be differentially private..

*3. Adding DP to machine learning model predictions:* When the model itself does not need to be published, DP can be applied to the model's predictions. Although this method is considered to be relatively weaker than the previous two. But if this method is combined with other privacy measures (e.g., only authorised users can access it), then it is also considered appropriate [38].

## 3 Literature review

Privacy can be approached from various perspectives, but since this paper focuses on privacy issues in LA applications, particularly those involving machine learning approaches, it is best aligned with the definition provided by [20]. According to [20]'s definition, privacy is the consent individuals give for data collection, with the expectation that the use and release of their information will not cause harm, though potential risks can arise if adversaries get access to their personal details from the data or the outputs [20].

### 3.1 Privacy Challenges in LA

From the technical perspective, privacy issues in LA may be broken down into three broad categories. First, the collection of sensitive and often large amounts of data [30]. With the increasing adoption of online courses and multimodal LA, the volume and variety of learning data also keep increasing. As a result, more sensitive data is being collected, extending beyond traditional demographic data to include student IP addresses, active times, affective states, social interaction data, and more. This necessitates further technical measures to ensure privacy protection [15]. Second, insufficient anonymization and computation of sensitive data [30]. As data volume increases, meeting the stringent requirements of data protection laws (such as GDPR), which require sensitive data to be anonymized to an unidentifiable level, has become a new challenge. Finally, advances in AI technologies have introduced new issues, particularly concerning potential adversarial attacks on learning-related data. For instance, [45] demonstrated that even de-identified student data can be easily exposed under unsupervised learning adversarial attacks.

---

[5]when a model is said "differentially private", it means the model satisfy the definition of DP

## 3.2 Previous privacy solutions in LA

The current privacy solutions in LA are insufficient to adequately address privacy challenges identified in this area as reported by [24]. The first type of solutions are framework-based solutions, which tend to be conceptual and majority of them lack empirical validation [30]. Moreover, while such approaches have their strengths, they tend to be less directly helpful for addressing privacy challenges mentioned above (Section 3.1). On the other hand, while some technical solutions demonstrate good results, they may be designed for a particular learning modality and not easily transferable across different contexts. For instance, the MOOC replication framework (MORF) proposed by [15] allows researchers to use data without direct access, but since it is designed specifically for the MOOC environment, it cannot be transferred to other contexts.

More technical privacy solutions in LA tend to rely on traditional anonymization methods, which offer transferability and allow anonymized data to be used in downstream data analysis tasks. For example, [28] employed an open-source software tool called ARX, which applied various anonymization techniques, including k-anonymity, to assess the potential risks of re-identification. Similarly, [42] and [41] utilised k-anonymity in an attempt to enhance privacy in educational data mining while maintaining balance in downstream machine learning tasks. However, k-anonymity and other anonymization methods commonly used for student data, are considered relatively outdated, and their use as privacy-enhancing technologies has been in decline since the emergence of DP [2]. This is because these traditional anonymization methods have been proven vulnerable to linkage attacks. A linkage attack happens when an attacker has access to information related to anonymized data (e.g., background knowledge), in which case they can make use of this information to re-identify individuals, leading to a breach of the anonymized dataset. A recent example is the Illuminate Education company that failed to protect personal information of approximately 820,000 current and former students from linkage attacks [22].

In this context, DP may offer robust protection against linkage attacks and stand against many other attacks on sensitive data, such as bias and similarity attacks [32]. This is crucial for preventing personal identity disclosure and privacy breaches, especially in the face of advanced AI technologies. Additionally, DP is more effective when handling large-scale datasets, as it can scale to large datasets and complex queries, whereas k-anonymity may become impractical as the dataset size increases [19]. Few LA researchers have started examining DP for dealing with LA privacy issues. One of the earliest examples is [13], who simultaneously used DP, k-anonymity, and l-diversity on educational tabular data, recording utility performance and advocating for the broader application of DP in LA tasks. [37] shared a similar conclusion, suggesting that DP could be embedded as an independent module within the privacy services of educational software products, where it could play a significant role. Building on this, [16] went a step further by applying both k-anonymity and DP to evaluate the privacy protection results on the same dataset, varying the DP privacy parameter $\epsilon$ to demonstrate the trade-off between utility and privacy. Additionally, recent studies have explored the combination of DP with other methods, such as synthetic data generation. Synthetic data offers a favourable trade-off between utility and privacy in educational datasets, maintaining near-realistic data levels of utility while enhancing privacy [31][25]. Therefore, this combination is considered highly promising [38]. Research by [47] [29]demonstrated that synthetic data with DP performs well in maintaining data utility compared to privacy-preserving methods like perturbation and binning, with Utility Loss gradually decreasing as $\epsilon$ increases.

The above suggests that the use of DP in LA is still in its infancy. While DP has shown great potential, there is still a lack of extensive research in this area. Therefore, we propose a framework aimed at guiding the LA community to apply DP throughout the LA process (Section 4). Additionally, we demonstrate the performance of the DP mechanism in protecting privacy while maintaining predictive accuracy, through a realistic LA usage scenario (Section 5). By offering detailed guidance based on a real-world educational dataset, we aim to overcome the current limitation of presenting DP, which is often explained through very few showcase examples [6].

## 4 DEFLA (Differential privacy framework for learning analytics) Framework

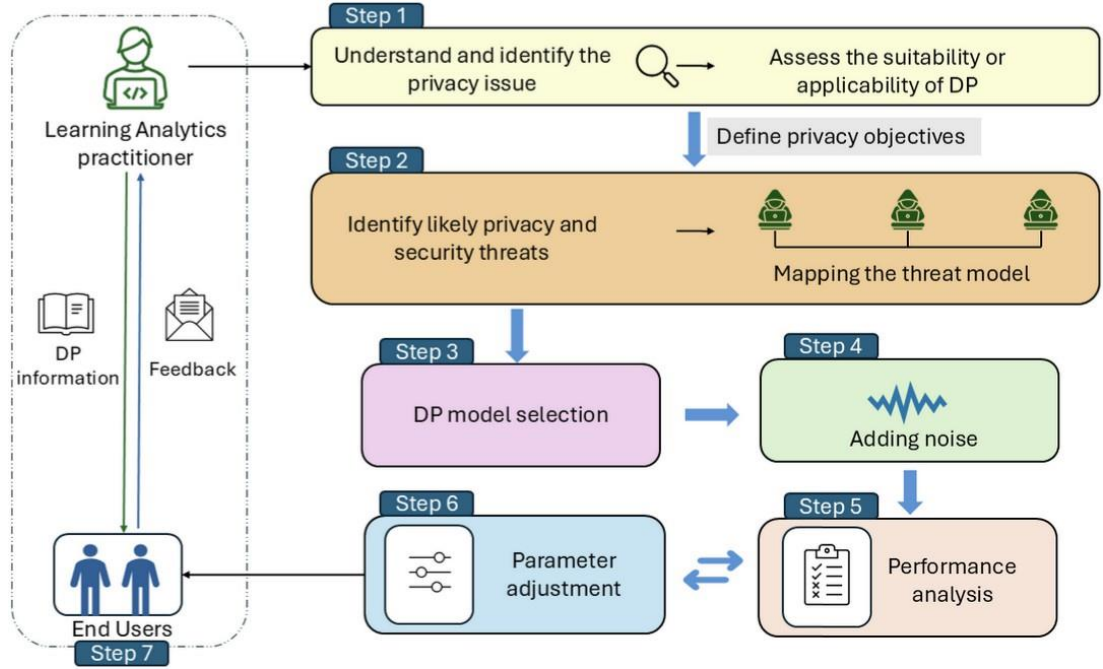

Fig. 1. DEFLA - the proposed learning analytics DP framework.

As our framework is aimed at enabling privacy-preserving learning analytics with DP, we name it DEFLA (Differential privacy framework for learning analytics) (Figure 1). It is derived from three main sources. The first source is suggestions originating from the prior research on implementing DP in LA. For example, [16] suggest, in their conclusion, that it's important to map the attacker model, specify the types of information collected in the dataset, and, as emphasised by many studies, maintain the balance between data privacy protection and information loss as much as possible [13] [47]. The second source is the report of an authoritative workshop on DP [6]. This workshop was held in 2022 at Harvard University, bringing together experts from industry, academia, and the public sector to discuss better ways to address the current challenges of deploying DP. The recommendations from [6]'s report on how to implement DP have informed a large part of this framework (including first to fourth step of DEFLA). The third source is a user expectation survey on DP conducted by [7]. The survey results, derived from over 2,000 responses, show that only informing end users that a system uses DP does not increase their willingness to share personal information. Users are concerned about the types of information leakage that DP protects against. After receiving a detailed explanation of DP, users might be more willing to share their private data with trusted parties [7]. This outcome forms the final step of DEFLA, which addresses user communication issues. Below, we introduce each step of DEFLA.

First, when LA practitioners encounter potential privacy issues, the **first step** is to understand the nature of the data privacy problem. Since there are many types of privacy issues that may arise in a LA process, some of these problems are not suitable for DP. In general, DP is suitable when working with data of high volume and complexity, sharing data with privacy requirements, and training a wide range of machine learning model [5]. However, DP becomes unsuitable for scenarios with smaller data scales, scenarios with high demand for precise analysis, scenarios without high privacy requirements, and scenarios requiring real-time data analysis or rapid response time [5]. Therefore, practitioners need to understand the specific privacy issue at hand, define the privacy objectives, and assess whether DP can provide a solution. Practitioners may refer to the book "Hands-on Differential Privacy" by [5] to learn more about distinct contexts where DP is applicable.

In the **second step**, once practitioners have identified their scenario as one of the three scenarios that DP is suitable for, they can begin to map out a threat model. A threat model refers to the most likely threats to privacy and security. Threat modelling identifies potential threats and develops countermeasures by answering questions such as "Where am I most vulnerable to attack?", "What are the most relevant threats?", and "What do I need to do to defend against these threats?" [9]. There are many commonly used threat modelling frameworks available for practitioners, such as the STRIDE framework developed by Microsoft and the seven-step PASTA threat analysis process [33][26]. In the academic research on applying DP to machine learning, there are two common threat models. One is designed to defend against a "worst-case scenario," assuming that the adversary knows the DP mechanism, has full access to the training database, and can extensively manipulate the machine learning model itself [21]. The other, proposed by [21], is a more realistic adversarial threat model called the relaxed threat model, where the adversary does not have full access to the training database, but may have related or partial information. Overall, by defining a threat model, LA practitioners can clarify their needs for a DP model in the next step.

The **third step** is the selection of the DP model and settings. This process is relatively complex and is recognized as one of the practical challenges of DP [6]. Choosing the appropriate DP model involves deciding at which stage to add noise and selecting the noise mechanism. Establishing a threat model (in Step 2) can help clarify which stakeholders are trusted; if the data collector is not trusted, noise can be added at the data input stage. Conversely, if the data collector is trusted, noise can be added during training or in the prediction results. Additionally, the structure and characteristics of the dataset are important considerations. For example, different noise mechanisms are suited for different types of data: the Laplace mechanism is often used for numerical data, the Gaussian mechanism for high-dimensional data or situations where less noise is needed, and the exponential mechanism for non-numerical data or when the output space is large and discrete [38]. If the data is to be analysed iteratively multiple times, then an interactive approach is preferred; otherwise, a non-interactive approach is more suitable. While this step covers the primary choices in DP model and settings, additional decisions (such as privacy budget) may be informed by the results of the performance analysis in the fifth step.

The **fourth step** involves applying DP to the dataset or data pipeline of the LA, based on the DP models and settings determined in the previous step. It is important to note that data preprocessing is crucial when deploying DP. Noise mechanisms such as Laplace and Gaussian are sensitive to the data range, therefore performing data normalisation is essential for practitioners who choose these noise mechanisms. Additionally, DP is sensitive to outliers, which can excessively amplify the impact of noise, thereby affecting the stability of the model [34].

The **fifth step** is performance analysis. Performance analysis includes both the evaluation of the performance of machine learning models applying DP, such as accuracy and F1-score, and the evaluation of privacy performance, i.e., privacy audit. The evaluation of DP machine learning models uses the same metrics as non-DP machine learning evaluations, thus not discussed here. Privacy auditing is performed by the system designer to detect privacy leakage in the system [6]. It is also commonly used in non-DP machine learning models to measure privacy leakage in machine learning models. Privacy auditing in DP machine learning involves two key components: (a) $\epsilon$-values, which provide theoretical privacy guarantees; (b) attacks, which help to validate and discover privacy vulnerabilities in the system and are used to optimise DP mechanisms. The value of $\epsilon$ in DP sets an upper bound on the level of privacy protection, indicating the maximum potential privacy leakage in the worst-case scenario. On the other hand, simulated attacks are used to evaluate the system's lower bound of privacy by testing possible attack scenarios to uncover any additional privacy risks [38]. Common privacy auditing methods include membership inference attacks (MIA) and data extraction attacks [6]. Overall, the combination of $\epsilon$-values and simulated attacks enables a comprehensive assessment of the effectiveness of privacy protection.

The **sixth step** is the adjustment of the DP parameters based on the results of the performance analysis in the previous step. These parameters primarily include the epsilon value, noise distribution mechanism, sensitivity, sampling rate (applicable in specific DP algorithms like PATE), binning and clipping, and the gradient clipping threshold (in DP for deep learning) [32]. Both the fifth and the sixth steps are iterative; after completing the sixth step, it is necessary to conduct another performance analysis. This process should continue until a satisfactory balance is achieved for the practitioner and all stakeholders involved in LA.

The **final step** is user communication. Research shows that end users are more willing to share personal information when they understand how DP protects against specific types of information disclosure, and they are informed that the risks are minimised [6]. Therefore, explaining the DP-based privacy protection mechanism thoroughly to end users is crucial for enhancing their trust in the LA system and obtaining their consent for data use. [6] suggest that the ideal approach to communicating DP to users is to ensure that every user contributing data knows how their data is being used, for what purposes, what level of DP is being implemented, and that they are allowed to verify these claims. Of course, user communication is not just about the LA practitioner conveying DP-related information to the users, but also gathering user feedback on the entire DP-based LA process, and iteratively improving the process.

It is essential to maintain transparency throughout the entire DP implementation process. Practitioner should monitor and document every detail of the DP implementation. Additionally, sharing necessary information with regulatory bodies and other stakeholders involved in the process, such as students and teachers, is an important aspect of ensuring transparency. The necessary information to be shared includes the scope of data protected by DP, the achieved outcomes, the risks mitigated, and legal compliance. If stakeholders express a desire for more detailed DP implementation information, it should be offered upon request. One example is Amazon's application of DP in its AWS Clean Rooms product, which provides a detailed explanation of how DP functions and assists customers in mitigating risks [1]. Amazon also offers video demonstrations and additional channels for interested users to learn more [1].

## 5  Puting framework in use: Validation of DP in LA setings

### 5.1  Experiment Setup

We demonstrate DEFLA through a LA experiment using the popular dataset from the Open University (OULAD). To that end, we assume a scenario based on this dataset and following the steps of the DEFLA, provide guidance on deploying DP step by step.

In this scenario, we assume that a LA practitioner at a university has obtained a real LA dataset (described below). The practitioner's goal is to train a machine learning model on this dataset to predict whether students will pass a particular course. After constructing the prediction model, the practitioner needs to release a de-identified version of the dataset, along with the machine learning model, and its prediction results. In the following, we first introduce the dataset, machine learning model, and experimental environments, and then proceed with the details of the experimental process using the DP framework.

**Dataset**. Open University Learning Analytics Dataset (OULAD) [28] originates from online courses at the Open University and includes data from seven courses, students enrolled in those courses, and the students' interactions with the Virtual Learning Environment. This dataset also contains various personal details about students, such as age and residence. The dataset is anonymized, with unique identifiers like social security numbers removed, numeric identifiers (e.g., *student_id*) randomised, and further anonymization applied using k-anonymity [28]. This dataset is particularly suitable for our experiments, as it allows for verifying whether traditional anonymization methods allow for protecting students' privacy in the current technological landscape, that is, whether more robust protection is needed. **Machine learning model**. For this experiment, we select logistic regression, a widely used model in LA. Following the approach of [18], the model was trained using $\ell_2$ regularisation with a regularisation parameter of $\lambda = 0.0001$ over 100 epochs. The dataset was initially split into two equal parts. The first half was further divided into training and test sets for the machine learning model, while the remaining half was allocated to training and test sets for the membership inference attack model. **Experimental environment**. The entire experiment is conducted on Google Colab, with 13 GB of RAM and 108 GB of disk space, using IBM's differential privacy package, Diffprivlib [14].

## 5.2 Experiment Procedure

**Step 1:** Based on the guidance from step 1 of the DEFLA, using DP for this specific LA scenario is highly appropriate. In particular, DP is highly suitable for machine learning predictions, as is the case in the current scenario. Additionally, the practitioner's privacy objective is to ensure compliance with relevant legal requirements on privacy, while also maintaining prediction accuracy as much as possible.

**Step 2:** Setting the threat (adversary) model. By building a threat model, practitioners will have a better understanding of the types and sources of potential threats, the level of privacy protection they require, and methods for verifying privacy. To accomplish the task of adversary model development, we suggest using the previously mentioned STRIDE framework for threat analysis [33]. To exemplify the use of STRIDE framework, we have created a threat model for our hypothetical scenario and made it available in the supplementary file[6]. In brief, the threats identified in this threat model include competitors and malicious actors attempting to reverse-engineer the model to gain access to the real training data, which contains personal information. Potential attacks, such as membership inference attacks, are also considered as a possible threat. However, it is important to note that, since this scenario is hypothetical, several details remain unspecified. For example, while most users in the dataset are based in the UK, there is no indication of the practitioner's location, whether cross-border data transfers are involved, which legal frameworks apply, or whether any additional stakeholder requirements exist. As such, this threat model should be viewed as a reference point. To facilitate a better understanding of DP implementation, in the next step (Step 3), we will simulate various DP models and configurations within the constraints of this simplified threat model.

**Step 3:** As mentioned in step 2, we simulate various DP models and configurations within the constraints of this simplified threat model. Therefore, our experiments address the three stages of applying DP within the machine learning pipeline (as outlined in Section 2.2): DP at the input level, during model training, and at the prediction stage. The methods we adopt for each stage are illustrated in Figure 2. Among these three stages, since applying DP earlier in the process typically provides stronger privacy protection [18], the first stage, which involves DP at the input level, is expected to offer the strongest privacy guarantees. The following are the details for each stage methods:

1. In the case of adding DP at the input/data level, we use the input perturbation method. Specifically, we use the input perturbation method from [12]. Here, the data collector is untrusted and cannot access raw learner data. In contrast, all subsequent methods assume a trusted data collector with access to the collected raw learner data. This method excels in maintaining low excess empirical risk[7] compared to other perturbation techniques like objective perturbation, effectively preserving privacy without significantly compromising data utility [12].

2. When adding DP during the machine learning training process, we use the objective perturbation technique proposed by [3], as this method also has good records in balancing privacy and utility [3].

3. For adding DP to machine learning predictions, we implement the PATE prediction perturbation method, proposed by [35]. It adds noise to the predictions before generating the final labels. This method is particularly effective for

---

large-scale and complex learning tasks due to its robust privacy protections and its capacity to efficiently manage uncurated, imbalanced datasets [35].

We cover a broad range of privacy budgets, with the following list of $\epsilon$ values: [0.01, 0.1, 1, 10, 100, 1000, 10000]. Based on recommendations from the literature, input perturbation uses Gaussian noise, while the objective perturbation and prediction perturbation use Laplace noise [36].

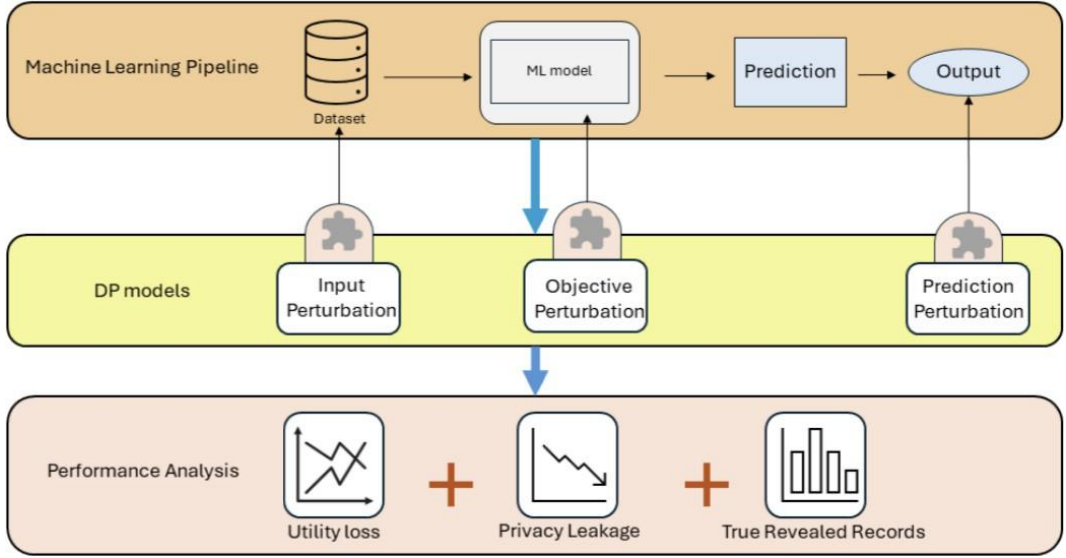

Fig. 2. illustrates Details of framework Step 3 (selection of DP models), Step 4 (applying DP) and Step 5 (performance analysis) illustrated through the experimental example.

**Step 4:** Applying DP. First, we preprocessed the dataset, including normalising the numeric features to ensure consistency in scale, which is important for DP mechanisms. Outliers were minimal, likely due to the k-anonymity processing applied before the dataset was published. Additionally, we transformed the original multi-class target variable into a binary format to facilitate logistic regression and applied OneHotEncoder (a technique that converts categorical data into a binary matrix) to the categorical variables, making the data suitable for subsequent machine learning tasks. After completing all the pre-processing, we applied the three DP methods mentioned in step 3 (i.e. input perturbation, objective perturbation and prediction pertuerbation) to the machine learning process.

**Step 5:** In the performance analysis step, as shown in Figure 2, we adopt three effective metrics for analysing the performance of DP in the context of machine learning predictive models, as proposed by [18]. These metrics are Utility Loss, Privacy Leakage, and True Revealed Records. Utility Loss represents the utility difference between non-private and differentially private models. When Utility Loss is 0, it indicates that the privacy model achieves the same utility as the non-private model. Formally, Utility Loss is calculated as:

$$\text{Utility Loss} = \text{accuracy of private model} - \text{accuracy of non-private model [18]}.$$

Privacy Leakage quantifies the difference between the true positive rate and false positive rate of the adversary's inference attack. True Positive Rate is the proportion of correctly identified positive cases (accurate inferences by the adversary), while False Positive Rate is the proportion of negative cases incorrectly classified as positive. Its value range is [0, 1]. A Privacy Leakage of 0 indicates that the inference attack does not lead to data leakage, while a value of 1 suggests the attack is fully successful. In some cases, Privacy Leakage may be negative, indicating that the attack model may have falsely detected non-members as members, implying poor performance by the attack model and effective DP protection. True Revealed Records estimates the actual number of members at risk of data leakage when the membership inference attack is successful.

**Step 6:** Adjusting Parameters Adjusting parameters based on performance analysis. According to the privacy objectives set in Step 1, we aim to maintain good performance of the machine learning model while achieving strong privacy protection. Therefore, in this scenario experiment, our goal is to strike a balance between utility and privacy, rather than prioritising one over the other. Under these privacy objectives, we seek a balance between Utility Loss and Privacy Leakage computed in step 5. Consequently, the experiment will report Utility Loss and Privacy Leakage across a wide range of privacy budgets (i.e., epsilon values). If no clear balance point is observed in the results, we will fine-tune the privacy budget parameters and report on the resulting Utility Loss and Privacy Leakage again, searching for a point with minimal trade-offs. Additionally, if significant anomalies arise, such as the trend of Utility Loss for the epsilon values diverging significantly from what is reported in the literature, we will re-examine factors such as the compatibility between the noise mechanism and the dataset.

**final step,** user communication, varies significantly depending on the specific context. For instance, the end-users for some LA practitioners might be K-12 students, while others might serve adult learners in vocational education. In this experiment, the dataset originates from the Open University, with the majority of users being adults. Therefore, we would adopt the following approach for communicating the deployment of DP to users: First, we would explain the principles of DP in simple language through the user interface in the privacy note page, highlighting the advantages of DP to alleviate possible concerns that end users may have about sharing personal information. For users who want to learn more, we would include a link to additional content that further explains DP and justifies our use of DP in the given learning settings, similar to what the US Census Bureau has done (US Census Bureau, 2020). For a more technical audience, we would include details about the DP deployment, such as the type of noise mechanism used, the stage at which noise is added, and the privacy budget. Each DP detail would be supplemented with an explanatory note to help end users understand its meaning. Finally, we would provide a channel for user feedback, as Microsoft did in their privacy tool, encouraging users to ask questions or provide feedback regarding data privacy [8]. The intention is to increase user engagement and build trust in the system.

## 5.3 Experiment Results

This section focuses on Step 5 of DEFLA, namely the evaluation of the performance of DP models across various privacy budget (Epsilon, $\epsilon$) values, highlighting the trade-off between utility and privacy. We examine how different DP models, applied in different stages of machine learning modelling (Section 5.2, Step 3), perform in terms of Utility Loss, Privacy Leakage, and True Revealed Records.

*5.3.1 Utility Loss analysis.* Figure 3 depicts Utility Loss as a function of privacy budget (epsilon) for different DP models. Among these, the input perturbation model shows the lowest Utility Loss compared to the other perturbation models. The prediction perturbation model, which initially shows the highest Utility Loss, eventually decreases to nearly zero as $\epsilon$ increases. In contrast, the objective perturbation model shows fluctuating Utility Loss but decreases and becomes stable after $\epsilon = 10$. Overall, objective perturbation is the worst choice compared to the other two models in terms of utility preservation when working with the chosen dataset. On the other hand, the Utility Loss is the lowest for input perturbation, which is contrary to previous research where the Utility Loss is the lowest when the noise is added later in the machine learning process [18]. This finding might be due to the particularities of this specific dataset.

*5.3.2 Privacy audit.* As expected, and consistent with the literature [18], [36], Figure 4 shows that the prediction perturbation model undergoes increasing Privacy Leakage as the privacy parameter $\epsilon$ increases. After $\epsilon = 1$, the leakage value stabilises at 0.02, indicating potential Privacy Leakage. As for the input perturbation, it has a good privacy protection effect, and its highest Privacy Leakage value is lower than the highest value of the other two methods. This is also consistent with the experimental results reported in the literature (e.g., [18]). After all, input perturbation directly adds noise at the input/data level and can achieve a strong level of privacy protection. A notable phenomenon is that the "input perturbation" line shows negative Privacy Leakage after $\epsilon = 0.1$, indicating that the attack model is more likely to misidentify non-members as members, a result of high false positive rates. Interestingly, this phenomenon of negative Privacy Leakage also occurs in the other two methods, suggesting that all approaches may exhibit a 'security through obscurity' effect, where the data appears to be protected because the attack model's predictions are often wrong, confusing attackers by incorrectly labelling outsiders as insiders [18]. The objective perturbation shows an initial Privacy Leakage of 0.02 and fluctuates around the zero level, which indicates that the Privacy Leakage is minimal across the range of $\epsilon$ values.

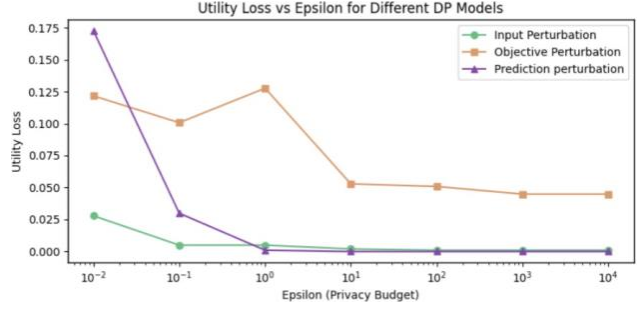

Fig. 3. Utility Loss for logistic regression model. The higher the value on the y-axis, the higher the data Utility Loss; the higher the epsilon value (x-axis), the less noise is added.

Figure 5 depicts the number of True Revealed Records for the DP models with respect to the privacy budget ($\epsilon$). As the privacy budget $\epsilon$ increases, the prediction perturbation model shows only a minimal and gradual increase in the number of true positive records, but this increase is very negligible, keeping the true records revealed by prediction perturbation close to zero across all values of $\epsilon$. While the objective perturbation model shows a stable count initially, the number of True Revealed Records increases as $\epsilon$ increases above one. The input perturbation model has no True Revealed Records at low $\epsilon$ but sharply increases and stabilises after $\epsilon = 10^{-1}$. Among the models, prediction perturbation shows the lowest number of true positive records.

Having completed step 5, the performance analysis, we now proceed to step 6, which is adjusting parameters based on performance analysis. As stated earlier in Step 6 section 5.2, our objective is to achieve a balance between utility and privacy, rather than prioritizing one over the other. Based on the results of the performance analysis, we found that the most balanced utility and privacy trade-off is achieved through prediction perturbation, particularly when $\epsilon = 10^{-1}$. Since a relatively clear balance point has already been identified, no further parameter adjustment will be conducted in this demonstration experiment.

## 6  Discussion and Limitations

First, the experiment highlights the significant role that DP plays in ensuring privacy protection, showing that DP can address privacy issues that traditional anonymization techniques fail to resolve. Furthermore, the privacy audit presented in this paper extends previous work on the application of DP in LA by covering adversarial techniques and reporting the true positive and false positive rates of attacks (i.e., Privacy Leakage analysis) as well as the data points at risk after a successful attack (True Revealed Records). This approach moves beyond prior assessments that rely solely on epsilon values, providing a clearer demonstration of DP's role in privacy protection. From the privacy audit conducted in the experiment, it can be observed that when the DP privacy parameter epsilon is larger, the risk of Privacy Leakage from membership inference attacks increases across all three DP methods. Similarly, the True Revealed Records follows the same trend – as DP's privacy protection weakens, the number of data points exposed by successful membership inference attacks increases. This indicates that when DP epsilon is large and privacy protection is minimal, even if the dataset has been anonymized (as was the case with the experimental dataset, see Section 5.1), it still faces significant privacy threats. This also shows that traditional anonymization methods (e.g., k-anonymity) are not robust against more advanced adversarial attacks. This aligns with the findings of [45], who concluded that traditional anonymization methods are becoming insufficient in the face of continuously evolving AI-driven attack methods, and that the LA field requires cutting-edge privacy-enhancing technologies such as DP. Additionally, considering the increasing size of datasets in the LA domain, traditional anonymization techniques struggle to meet the de-identification requirements set by data protection laws [30]. Our experiments, which demonstrated strong privacy protection on relatively large datasets (sample size > 30,000), suggest that DP is also highly effective for handling such large datasets. Finally, while the experimental dataset used in this study was limited to tabular data, DP has also performed well on image and text data, showing good extensibility [27, 46]. The extensibility of DP across different data modalities can be leveraged to handle the growing use of multimodal data in the LA domain.

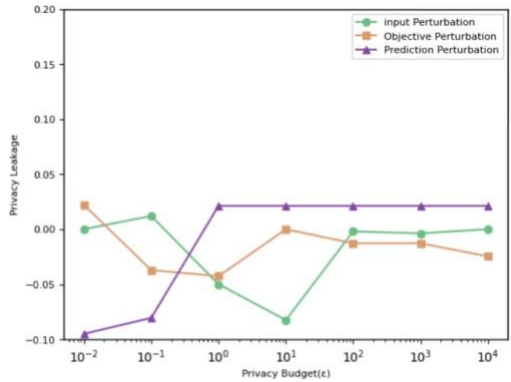 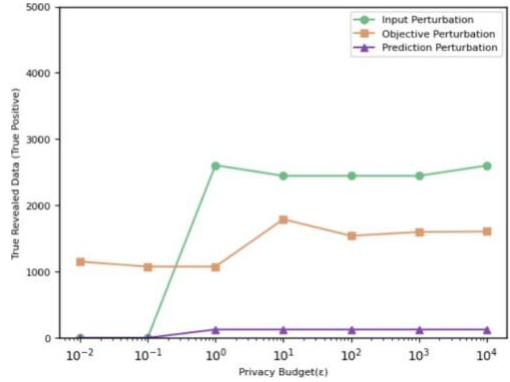

Fig. 4. Privacy Leakage for the examined DP models. Higher values on the y-axis mean higher likelihood of Privacy Leakage; higher values of epsilon mean less noise is added.

Fig. 5. True Revealed Records for the examined DP models. Higher values on the y-axis mean more true records are revealed, whereas higher epsilon values mean less noise is added.

Second, the experiments clearly demonstrate that any privacy-enhancing technique, including DP, inevitably involves a trade-off between privacy and utility. From the Utility Loss analysis (Figure 3) and the privacy audit results (Figure 4), it is evident that as epsilon increases, Utility Loss gradually decreases, but Privacy Leakage increases. Moreover, if a particular DP method exhibits better Utility Loss performance, its results on the privacy audit are not as favourable. For example, input perturbation shows the lowest Utility Loss among the three methods, but the highest True Revealed Records. These findings also validate the critique by [40] regarding technical privacy-preserving methods, which always require careful consideration of the trade-off between privacy and utility. This outcome suggests that the community should continue to strive for the development and optimization of privacy-enhancing techniques and explore the application of non-technical measures to compensate for the utility-privacy trade-off limitations of technical approaches.

Finally, we proposed a framework called "DEFLA" to help the LA community apply DP, and demonstrated its use in scenarios based on a well-known LA dataset. This addresses the shortcomings of many frameworks aimed at solving privacy issues in LA, which often lack practical evidence [30]. Demonstrating DEFLA with a LA dataset and a realistic LA usage scenario not only enhances its credibility but also provides practitioners interested in using DEFLA with a high fidelity example. This is especially important considering the current lack of examples on how to apply DP in specific application settings [6].

The limitation of this paper is as follows: Although DP is valued for its relatively low computational complexity and suitability for machine learning [32], challenges persist due to its computational demands. For example, DP methods like DP-SGD, widely used in deep learning, require gradient clipping, which can hinder hardware acceleration on GPUs or TPUs [6]. In this study, executing some DP methods on a dataset of over 30K rows took several hours, reflecting their computational intensity. Additionally, experiments were conducted outside deep learning contexts. DP also struggles with small datasets (N<50), common in LA, where its performance can degrade significantly [34]. Further research is needed to explore DP's performance across diverse dataset sizes and LA scenarios. Finally, while DEFLA was tested in a basic LA setup, its effectiveness in more complex, real-world LA applications requires validation to ensure broader credibility.

## 7  Conclusions

In this paper, we present the first practical framework for implementing DP in the LA domain, DEFLA. DEFLA is informed by insights from the literature on DP implementation in LA, the characteristics of the LA process, and the recommendations of the DP community for DP implementation. DEFLA addresses the lack of comprehensive and systematically defined guidelines for applying DP in LA. By offering a well-defined and clearly structured practical procedure for DP use, this framework lays the foundation for a wide adoption of DP within the LA applications. Furthermore, our results demonstrate the significant role of DP in privacy preservation in the context of a realistic LA usage scenario and dataset. We showed that even though the privacy-utility trade-off is present in the dataset, the Utility Loss is ideal (i.e., close to zero) even when the privacy guarantee is very conservative (i.e., epsilon = 0.1).

The implications for the LA field from this study are multifold: 1) The experiments demonstrate that traditional anonymization methods are no longer sufficient against advanced attacks. DP, known for robust protection against both current and future threats [30], should be more widely adopted in LA, especially for machine learning-based analyses. 2) The adoption of DP in the LA field is still limited, and its use can be challenging. DEFLA presented in this paper provides practical steps and a systematic procedure for applying DP in LA, helping practitioners effectively implement this advanced privacy-enhancing technology, ensuring compliance and security in data processing. 3) By examining the effects of three different DP methods on privacy protection and Utility Loss, the paper thoroughly analyses the balancing of privacy and utility and offers relevant insights into DP's different options for achieving this balance. This valuable practical experience offers a foundation for further exploration and optimization of DP in LA to enhance trust in data use within the field.

### Acknowledgments

This research is co-funded by the Trond Mohn Research Foundation, project "EduTrust", nr. TMS2023TMT03